\documentstyle[12pt,fleqn]{article}

\font\twlgot =eufm10 scaled \magstep1 \font\egtgot =eufm8
\font\sevgot =eufm7 \font\twlmsb =msbm10 scaled \magstep1
\font\egtmsb =msbm8 \font\sevmsb =msbm7

\newfam\gotfam
\def\pgot{\fam\gotfam\twlgot}
\textfont\gotfam\twlgot \scriptfont\gotfam\egtgot
\scriptscriptfont\gotfam\sevgot
\def\got{\protect\pgot}
\newfam\msbfam
\textfont\msbfam\twlmsb \scriptfont\msbfam\egtmsb
\scriptscriptfont\msbfam\sevmsb

\def\pBbb{\relax\ifmmode\expandafter\Bb\else\typeout{You cann't use
Bbb in text mode}\fi}
\def\Bb #1{{\fam\msbfam\relax#1}}

\def\thebibliography#1{\section*{References}\list
    {[\arabic{enumi}]}{\settowidth\labelwidth{#1}\leftmargin\labelwidth
      \advance\leftmargin\labelsep
      \usecounter{enumi}}
      \def\newblock{\hskip .11em plus .33em minus .07em}
      \sloppy\clubpenalty4000\widowpenalty4000
      \sfcode`\.=1000\relax}

\let\Large=\large

\def\op#1{\mathop{\fam0 #1}\limits}

\newcommand{\Ker}{{\rm Ker\,}}
\newcommand{\nm}[1]{\mid {#1}\mid}
\newcommand{\beq}{\begin{equation}}
\newcommand{\eeq}{\end{equation}}
\newcommand{\ben}{\begin{eqnarray}}
\newcommand{\een}{\end{eqnarray}}
\newcommand{\be}{\begin{eqnarray*}}
\newcommand{\ee}{\end{eqnarray*}}
\newcommand{\bea}{\begin{eqalph}}
\newcommand{\eea}{\end{eqalph}}

\newcommand{\cR}{{\cal R}}
\newcommand{\cL}{{\cal L}}

\newcommand{\cF}{{\cal F}}
\newcommand{\cS}{{\cal S}}

\newcommand{\bL}{{\bf L}}

\newcommand{\cG}{{\got g}}

\newcommand{\al}{\alpha}

\newcommand{\bt}{\beta}
\newcommand{\dl}{\delta}
\newcommand{\la}{\lambda}

\newcommand{\m}{\mu}
\newcommand{\n}{\nu}

\newcommand{\G}{\Gamma}

\newcommand{\vt}{\vartheta}

\newcommand{\si}{\sigma}

\newcommand{\w}{\wedge}

\newcommand{\wh}{\widehat}
\newcommand{\ol}{\overline}
\newcommand{\dr}{\partial}
\newcommand{\ar}{\op\longrightarrow}
\newcommand{\ot}{\otimes}

\newcommand{\ve}{\varepsilon}

\let\ssection=\section
\renewcommand{\section}{\setcounter{equation}{0}\ssection}

\newcounter{eqalph}
\newcounter{equationa}
\newcounter{remark}
\newcounter{example}
\newcounter{theorem}
\newcounter{proposition}
\newcounter{lemma}
\newcounter{corollary}
\newcounter{definition}
\newenvironment{eqalph}{\stepcounter{equation}
\setcounter{equationa}{\value{equation}} \setcounter{equation}{0}

\begin{eqnarray}}{\end{eqnarray}\setcounter{equation}{\value{equationa}}}

\def\theremark{\arabic{remark}}
\def\therexample{\arabic{remark}}

\def\thedefinition{\arabic{definition}}

\newenvironment{theo}{\refstepcounter{definition} \medskip
\noindent{\it Theorem \thedefinition:}}{\medskip}

\newcommand{\mar}[1]{}

\begin{document}
\hbox{}

\begin{center}

{\Large \bf BV QUANTIZATION OF A GENERIC DEGENERATE QUADRATIC
LAGRANGIAN}

\bigskip

{\sc D. BASHKIROV}
\medskip

{\small

Department of Theoretical Physics, Moscow State University, 117234
Moscow, Russia

E-mail:  bashkir@phys.msu.ru

}
\end{center}

\bigskip

{\small

\noindent Generalizing the Yang--Mills gauge theory, we provide
the BV quantization of a field model with a generic almost-regular
quadratic Lagrangian by use of the fact that the configuration
space of such a field model is split into the gauge-invariant and
gauge-fixing parts.



}

\bigskip
\bigskip

The Batalin--Vilkoviski (henceforth BV ) quantization
\cite{bat,gom} provides the universal scheme of quantization of
gauge-invariant Lagrangian field theories. Given a classical
Lagrangian, BV quantization enables one to obtain a gauged-fixed
BRST invariant Lagrangian in the generating functional of
perturbation quantum field theory. However, the BV quantization
scheme does not automatically provide the path-integral measure,
unless a gauge model is irreducible. We apply this scheme to a
generic degenerate (almost-regular) quadratic Lagrangian.

We follow the geometric formulation of classical field theory
where classical fields are represented by sections of fiber
bundles. Let $Y\to X$ be a smooth fiber bundle provided with
bundle coordinates $(x^\m,y^i)$. The configuration space of a
first order Lagrangian field theory on $Y$ is the first order jet
manifold $J^1Y$ of $Y$ equipped with the adapted coordinates
$(x^\m,y^i,y^i_\m)$, where $y^i_\m$ are coordinates of derivatives
of fields \cite{book}. A first-order Lagrangian is defined as a
density \mar{cmp1}\beq L=\cL dx: J^1Y\to\op\w^nT^*X, \qquad n=\dim
X, \label{cmp1} \eeq on $J^1Y$. The corresponding Euler--Lagrange
equations are given by the subset \mar{b327}\beq \dl_i\cL=(\dr_i-
d_\la\dr^\la_i)\cL=0,   \qquad
 d_\la=\dr_\la + y^i_\la\dr_i + y^i_{\la\m}\dr^\m_i, \label{b327}
\eeq of the second-order jet manifold $J^2Y$ of $Y$ coordinated by
$(x^\m,y^i,y^i_\la, y^i_{\la\m})$. Any Lagrangian $L$ (\ref{cmp1})
yields the Legendre map \mar{m3}\beq \wh L: J^1Y\ar_Y\Pi, \qquad
p^\la_i\circ\wh L=\dr^\la_i\cL, \label{m3} \eeq of the
configuration space $J^1Y$ to the momentum phase space
\mar{00}\beq \Pi=\op\w^n T^*X\op\ot_Y V^*Y\op\ot_Y TX\to Y,
\label{00} \eeq called the Legendre bundle and equipped with the
holonomic bundle coordinates $(x^\la,y^i,p^\m_i)$. By $VY$ and
$V^*Y$ are denoted the vertical tangent and cotangent bundles of
$Y\to X$, respectively.

Let us consider a quadratic Lagrangian \mar{N12}\beq \cL=\frac12
a^{\la\m}_{ij} y^i_\la y^j_\m + b^\la_i y^i_\la + c, \label{N12}
\eeq where $a$, $b$ and $c$ are local functions on $Y$. This
property is coordinate-independent since $J^1Y\to Y$ is an affine
bundle modelled over the vector bundle $T^*X\op\ot_Y VY$. The key
point is that, if a Lagrangian $L$ (\ref{N12}) is almost-regular,
it can be brought into the Yang--Mills type form as follows.

Given $\cL$ (\ref{N12}), the associated Legendre map (\ref{m3})
reads \mar{N13}\beq p^\la_i\circ\wh L= a^{\la\m}_{ij} y^j_\m
+b^\la_i. \label{N13} \eeq Let a Lagrangian $\cL$ (\ref{N12}) be
almost-regular, i.e., the matrix function $a$ is a linear bundle
morphism \mar{m38}\beq a: T^*X\op\ot_Y VY\to \Pi, \qquad
p^\la_i=a^{\la\m}_{ij} \ol y^j_\m, \label{m38} \eeq of constant
rank, where $(x^\la,y^i,\ol y^i_\la)$ are holonomic bundle
coordinates on $T^*X\op\ot_Y VY$. Then the image $N_L$ of $\wh L$
(\ref{N13}) is an affine subbundle of the Legendre bundle
(\ref{00}). Hence, $N_L\to Y$ has a global section. For the sake
of simplicity, let us assume that it is the canonical zero section
$\wh 0(Y)$ of $\Pi\to Y$. The kernel of the Legendre map
(\ref{N13})  is also an affine subbundle of the affine jet bundle
$J^1Y\to Y$. Therefore, it admits a global section \mar{N16}\beq
\G: Y\to \Ker\wh L\subset J^1Y, \qquad a^{\la\m}_{ij}\G^j_\m +
b^\la_i =0,  \label{N16} \eeq which is a connection on $Y\to X$.
With such a connection, the Lagrangian (\ref{N12}) is brought into
the form \mar{y47}\beq \cL=\frac12 a^{\la\m}_{ij} (y^i_\la
-\G^i_\la)(y^j_\m-\G^j_\m) +c'. \label{y47} \eeq

Let us refer to the following theorems \cite{book,jpa99}.

\begin{theo}\label{04.2}  There exists a linear bundle
morphism \mar{N17}\beq \si: \Pi\op\to_Y T^*X\op\otimes_YVY, \qquad
\ol y^i_\la\circ\si =\si^{ij}_{\la\m}p^\m_j, \label{N17} \eeq such
that \mar{N45}\beq a\circ\si\circ a=a, \qquad
a^{\la\mu}_{ij}\si^{jk}_{\mu\al}a^{\al\nu}_{kb}=a^{\la\nu}_{ib}.
\label{N45} \eeq
\end{theo}

The equalities (\ref{N16}) and (\ref{N45}) give the relation
$(a\circ\si_0)^{\la j}_{i\m} b^\m_j=b^\la_i$. Note that the
morphism $\si$ (\ref{N17}) is not unique, but it falls into the
sum $\si=\si_0+\si_1$ such that \mar{N21}\beq \si_0\circ a\circ
\si_0=\si_0, \qquad a\circ\si_1=\si_1\circ a=0, \label{N21} \eeq
where $\si_0$ is uniquely defined.

\begin{theo} \label{m15} \mar{m15}
There is the splitting \mar{N18}\ben && J^1Y=\Ker\wh
L\op\oplus_Y{\rm Im}(\si_0\circ
\wh L), \label{N18} \\
&& y^i_\la=\cS^i_\la+\cF^i_\la= [y^i_\la -\si_0{}^{ik}_{\la\al}
(a^{\al\m}_{kj}y^j_\m + b^\al_k)]+ [\si_0{}^{ik}_{\la\al}
(a^{\al\m}_{kj}y^j_\m + b^\al_k)]. \label{N18'} \een
\end{theo}

The relations (\ref{N21}) lead to the equalities \mar{m25}\beq
\si_0{}^{jk}_{\m\al}\cR^\al_k=0, \qquad \cF^i_\m=(\si_0\circ
a)^{i\la}_{\m j}(y^j_\la-\G^j_\la).  \label{m25} \eeq By virtue of
these equalities the Lagrangian (\ref{N12}) takes the Yang--Mills
type form \mar{cmp31}\beq \cL=\frac12
a^{\la\m}_{ij}\cF^i_\la\cF^j_\m +c'. \label{cmp31} \eeq

Indeed, let us consider gauge theory of principal connections on a
principal bundle $P\to X$ with a structure Lie group $G$.
Principal connections on $P\to X$ are represented by sections of
the affine bundle $C=J^1P/G\to X$, modelled over the vector bundle
$T^*X\ot V_GP$ \cite{book}. Here, $V_GP=VP/G$ is the fiber bundle
in Lie algebras $\cG$ of the group $G$. Given the basis
$\{\ve_r\}$ for $\cG$, we obtain the local fiber bases $\{e_r\}$
for $V_GP$. The connection bundle $C$ is coordinated by
$(x^\m,a^r_\m)$ such that, written relative to these coordinates,
sections $A=A^r_\m dx^\m\ot e_r$ of $C\to X$ are the familiar
local connection one-forms, regarded as gauge potentials. The
configuration space of gauge theory is the jet manifold $J^1C$
equipped with the coordinates $(x^\la,a^m_\la,a^m_{\m\la})$. It
admits the canonical splitting \mar{N31'}\beq
a^r_{\m\la}=\cS^r_{\m\la}+\cF^r_{\m\la}=
\frac12(a^r_{\m\la}+a^r_{\la\m}-c^r_{pq}a^p_\m a^q_\la)
+\frac12(a^r_{\m\la}-a^r_{\la\m} +c^r_{pq}a^p_\m a^q_\la)
\label{N31'} \eeq (cf. (\ref{N18'})), where $\cF$  is the strength
of gauge fields up to the factor 1/2. The Yang--Mills Lagrangian
on the configuration space $J^1C$ reads \mar{5.1'}\beq L_{\rm
YM}=a^G_{pq}g^{\la\m}g^{\bt\n}\cF^p_{\la
\beta}\cF^q_{\m\n}\sqrt{\nm g}\, dx, \qquad  g=\det(g_{\m\nu}),
\label{5.1'} \eeq where  $a^G$ is a non-degenerate $G$-invariant
metric in the dual of the Lie algebra of ${\got g}$ and $g$ is a
non-degenerate metric on $X$.

If the Lagrangian (\ref{cmp31}) possesses no gauge symmetries, its
quantization in the framework of perturbation quantum field theory
can be given by the generating functional \be
Z=N^{-1}\int\exp\{\int (\cL +iJ_i y^i)dx\}\op\prod_x[dy(x)] \ee of
Euclidean Green functions.

Let us suppose that the Lagrangian $L$ (\ref{cmp31}) is invariant
under some gauge group $G_X$ of vertical automorphisms of the
fiber bundle $Y\to X$ which acts freely on the space of sections
of $Y\to X$. Its infinitesimal generators are represented by
vertical vector fields $u=u^i(x^\m,y^j)\dr_i$ on $Y\to X$ which
give rise to the vector fields \mar{m47}\beq J^1u=u^i\dr_i +d_\la
u^i\dr_i^\la, \qquad d_\la=\dr_\la +y_\la^i\dr_i, \label{m47} \eeq
on $J^1Y$. Let us also assume that $G_X$ is indexed by $m$
parameter functions $\xi^r(x)$ such that
$u=u^i(x^\la,y^j,\xi^r)\dr_i$, where \mar{m48}\beq
u^i(x^\la,y^j,\xi^r)=u_r^i(x^\la,y^j)\xi^r
+u_r^{i\m}(x^\la,y^j)\dr_\m\xi^r \label{m48} \eeq are linear first
order differential operators on the space of parameters
$\xi^r(x)$. The vector fields $u(\xi^r)$ must satisfy the
commutation relations \be
[u(\xi^q),u(\xi'^p)]=u(c^r_{pq}\xi'^p\xi^q), \ee where $c^r_{pq}$
are structure constants. The Lagrangian $L$ (\ref{cmp31}) is
gauge-invariant iff its Lie derivative $\bL_{J^1u}L$ along the
vector fields (\ref{m47}) vanishes, i.e., \mar{m49}\beq (u^i\dr_i
+d_\la u^i\dr_i^\la)\cL=0. \label{m49} \eeq

In order to study the invariance condition (\ref{m49}), let us
consider the Lagrangian (\ref{N12}) written in the form
(\ref{y47}). Since \mar{y50}\beq
J^1u(y^i_\la-\G^i_\la)=\dr_ku^i(y^k_\la-\G^k_\la), \label{y50}
\eeq one easily obtains from the equality (\ref{m49}) that
\mar{y45}\beq u^k\dr_k a^{\la\m}_{ij}+ \dr_iu^k a^{\la\m}_{kj} +
a^{\la\m}_{ik}\dr_ju^k =0. \label{y45} \eeq It follows that the
summands of the Lagrangian $L$ (\ref{y47}) and, consequently, the
summands of the Lagrangian (\ref{cmp31}) are separately
gauge-invariant, i.e., \mar{y31}\beq
J^1u(a^{\la\m}_{ij}\cF^i_\la\cF^j_\m)=0, \qquad
J^1u(c')=u^k\dr_kc'=0. \label{y31} \eeq The equalities
(\ref{m25}), (\ref{y50}) and (\ref{y45}) give the transformation
law \mar{b1}\beq J^1u(a^{\la\m}_{ij}\cF^j_\m)=-\dr_i u^k
a^{\la\m}_{kj}\cF^j_\m. \label{b1} \eeq The  relations (\ref{N21})
and (\ref{y45}) lead to the equality \mar{y53}\beq
a^{\la\mu}_{ij}[u^k\dr_k\si_0{}^{jn}_{\mu\al} -\dr_ku^j
\si_0{}^{kn}_{\mu\al} - \si_0{}^{jk}_{\mu\al}\dr_ku^n
]a^{\al\nu}_{nb}=0. \label{y53} \eeq

For the sake of simplicity, let us assume that the gauge group
$G_X$ preserves the splitting (\ref{N18}), i.e., its infinitesimal
generators $u$ obey the condition \mar{y49'}\beq
u^k\dr_k(\si_0{}^{im}_{\la\nu}a^{\nu\m}_{mj})+
\si_0{}^{im}_{\la\nu}a^{\nu\m}_{mk}\dr_ju^k -
\dr_ku^i\si_0{}^{km}_{\la\nu}a^{\nu\m}_{mj} =0. \label{y49'} \eeq
The relations (\ref{y50}) and (\ref{y49'}) lead to the
transformation law \mar{m53}\beq J^1u(\cF^i_\m)=\dr_ju^i\cF^j_\m.
\label{m53} \eeq Since $\cS^i_\la=y^i_\la-\cF^i_\la$, one can
easily derive from the formula (\ref{m53}) the transformation law
\mar{m54}\beq J^1u(\cS^i_\m)=d_\la u^i-\dr_ju^i\cF^j_\la= d_\la
u^i-\dr_ju^i(y^j_\la-\cS^j_\la) =\dr_\la u^i +\dr_ju^i\cS^j_\la
\label{m54} \eeq of $\cS$. A glance at this expression shows that
the gauge group $G_X$ acts freely on the space of sections
$\cS(x)$ of the fiber bundle $\Ker\wh L\to Y$ in the splitting
(\ref{N18'}). Then some combinations $b^r{}_i^\m\cS^i_\m$ of
$\cS^i_\m$ can be used as the gauge-fixing condition \mar{y1}\beq
b^r{}_i^\m\cS^i_\m(x)=\al^r(x), \label{y1} \eeq similar to the
generalized Lorentz gauge in Yang--Mills gauge theory.

Turn now to the BV quantization of a Lagrangian system with the
gauge-invariant Lagrangian $\cL$ (\ref{cmp31}). We follow the
quantization procedure in \cite{bat,gom} reformulated in the jet
terms \cite{barn,bran01}. Note that odd fields $C^r$ can be
introduced as the basis for a graded manifold determined by the
dual $E^*$ of a vector space $E\to X$ coordinated by
$(x^\la,e^r)$. Then the $k$-order jets $C^r_{\la_k\ldots\la_1}$
are defined as the basis for a graded manifold determined by the
dual of the $k$-order jet bundle $J^kE\to X$, which is a vector
bundle \cite{book00,mpl}. The BV quantization procedure falls into
the two steps. At first, one obtains a proper solution of the
classical master equation and, afterwards, the gauge-fixed BRST
invariant Lagrangian is constructed.

Let the number $m$ of parameters of the gauge group $G_X$ do not
exceed the fiber dimension of $\Ker\wh L\to Y$. Then we can follow
the standard BV procedure for irreducible gauge theories in
\cite{gom}.

Firstly, one should introduce odd ghosts $C^r$ of ghost number 1
together with odd antifields $y^*_i$ of ghost number $-1$ and even
antifields $C^*_r$ of ghost number $-2$. Then a proper solution of
the classical master equation reads \mar{y0}\beq \cL_{\rm PS}=\cL
+ y^*_iu^i_C-\frac12c^r_{pq}C^*_rC^pC^q, \label{y0} \eeq where
$u_C$ is the vector field \mar{y7}\beq u_C=u_r^i(x^\la,y^j)C^r
+u_r^{i\m}(x^\la,y^j)C^r_\m \label{y7} \eeq obtained from the
vector field (\ref{m48}) by replacement of parameter functions
$\xi^r$ and its derivatives $\dr_\m\xi^r$ with the ghosts $C^r$
and their jets $C^r_\m$.

Secondly, one introduces the gauge-fixing density depending on
fields $y^i$, ghosts $C^r$ and additional auxiliary fields, which
are odd fields $\ol C_r$ of ghost number $-1$ and even fields
$B_r$ of zero ghost number. Passing to the Euclidean space-time,
this gauge-fixing density reads \mar{y2}\beq \Psi=\ol
C_p(\frac{i}{2}  h^{pr}B_r +b^p{}_i^\m\cS^i_\m), \eeq where
$h^{pr}(x)$ is a non-degenerate positive-definite matrix function
on $X$ and $b^p{}_i^\m\cS^i_\m$ are gauge-fixing combinations
(\ref{y1}).

Thirdly, the desired gauge-fixing Lagrangian $\cL_{\rm GF}$ is
derived from the extended Lagrangian \be \cL'_{\rm PS}=\cL_{\rm
PS}+i \ol C^{*p} B_p, \ee where $\ol C^{*p}$ are antifields of
auxiliary fields $\ol C_p$, by replacement of antifields with the
variational derivatives \mar{y12}\beq y^*_i=\frac{\dl \Psi}{\dl
y^i}, \qquad C^*_p=\frac{\dl \Psi}{\dl C^p}=0, \qquad \ol
C^{*p}=\frac{\dl\Psi}{\dl \ol C_p}= \frac{i}{2}  h^{pr}B_r
-b^p{}_i^\m\cS^i_\m \label{y12} \eeq (see  the formula
(\ref{b327})). We obtain \mar{y4}\beq \cL_{\rm GF}=\cL+\dl_i\Psi
u^i_C - B_p(\frac12  h^{pr}B_r -ib^p{}_i^\m\cS^i_\m). \label{y4}
\eeq Let us bring its second term into the form \be &&
(\dr_i\Psi-d_\la\dr_i^\la\Psi) u^i_C= \dr_i\Psi u^i_C+
\dr_i^\la\Psi d_\la(u^i_C)-
d_\la(\dr_i^\la\Psi u^i_C)=\\
&& \qquad J^1u_C(\Psi)-d_\la(\dr_i^\la\Psi u^i_C), \ee where
\mar{y6}\beq J^1u_C=u^i_C\dr_i +d_\la u^i_C\dr^\la_i, \qquad
d_\la=\dr_\la + y^i_\la\dr_i +C^r_\la \frac{\dr}{\dr C^r}
\label{y6} \eeq is the jet prolongation of the vector field $u_C$
(\ref{y7}). In view of the transformation law (\ref{m54}), we have
\mar{y8}\ben && J^1u_C(\Psi)=-\ol C_pb^p{}_i^\la
J^1u_C(\cS^i_\la)= -\ol C_pb^p{}_i^\la[\dr_\la u^i_rC^r +
u^i_rC^r_\la +\dr_\la u^{i\m}_r C^r_\m +u^{i\m}_r C^r_{\la\m}
+\nonumber\\
&& \qquad (\dr_ju^i_rC^r +\dr_ju^{i\m}_rC^r_\m)\cS^j_\la]= -\ol
C_p M^p_r C^r, \label{y8} \een where $M^p_r C^r$ is a second order
differential operator on ghosts $C^r$. Then the gauge-fixing
Lagrangian (\ref{y4}) up to a divergence term takes the form
\mar{y9}\beq \cL_{\rm GF}=\cL -\ol C_p M^p_r C^r - \frac12
h^{pr}B_pB_r+ iB_p b^p{}_i^\m\cS^i_\m. \label{y9} \eeq

Finally, one can write the generating functional \be &&
Z=N^{-1}\int \exp\{\int (\cL -\ol C_p M^p_r C^r - \frac12
h^{pr}B_pB_r
+i B_p b^p{}_i^\m\cS^i_\m +iJ_ky^k)dx\}\\
&& \qquad \op\prod_x [dB_p][d\ol C][dC][dy] \ee of Euclidean Green
functions. Integrating it as a Guassian integral with respect to
the variables $B_p$, we obtain \mar{y10}\beq Z=N'^{-1}\int
\exp\{\int (\cL -\ol C_p M^p_r C^r - \frac12 h_{pr}^{-1}
b^p{}_i^\m b^r{}_j^\nu\cS^i_\m \cS^j_\nu +iJ_ky^k)dx\}\op\prod_x
[d\ol C][dC][dy].  \label{y10} \eeq Of course, the Lagrangian
\mar{y11}\beq \cL -\ol C_p M^p_r C^r - \frac12 h_{pr}^{-1}
b^p{}_i^\m b^r{}_j^\nu\cS^i_\m \cS^j_\nu \label{y11} \eeq in the
generating functional (\ref{y10}) is not gauge-invariant, but it
is invariant under the BRST transformation \mar{m60}\ben &&
\vt=u^i_C\dr_i +d_\la u^i_C\dr_i^\la +\ol v_r\frac{\dr}{\dr\ol
C_r} +
 v^r \frac{\dr}{\dr C^r}  +d_\la v^r \frac{\dr}{\dr C^r_\la}
+d_\m d_\la v^r \frac{\dr}{\dr C^r_{\m\la}},
 \label{m60}\\
&& d_\la=\dr_\la +y^i_\la\dr_i + y^i_{\la\m}\dr_i^\m + C^r_\la
\frac{\dr}{\dr C^r} +C_{\la\m}^r\frac{\dr}{\dr C^r_\m}, \nonumber
\een whose components $v$ are given by the antibrackets \be
v^r=(C^r,\cL'_{\rm PS})=\frac{\dl \cL'_{\rm PS}}{\dl
C^*_r}=-\frac12 c^r_{pq} C^pC^q, \qquad
 \ol v_r=(C^r,\cL'_{\rm PS})=\frac{\dl \cL'_{\rm PS}}{\dl \ol C^{*r}}=iB_r
\ee restricted to the shell (\ref{y12}) and to the solution
$B_r=ih^{-1}_{rp} b^p{}_i^\m\cS^i_\m$ of the Euler--Lagrange
equations $\dl\cL_{\rm GF}/\dl B_r=0$.

For instance, the generating functional (\ref{y10}) in the case of
$\cS^r_{\m\la}$ (\ref{N31'}), $h_{pr}=a^G_{pr}$ and
$b^p{}_r^{\nu\m}=\dl^p_rg^{\nu\m}$ restarts the familiar BV
quantization of the Yang--Mills gauge theory.

\end{document}